\documentstyle{elsart}

\begin{document}

\begin{frontmatter}
\title{Resonances in Fock Space: Optimization of a SASER device}

\author [Arg] {L.E.F. Foa Torres},
\author [Arg] {H.M. Pastawski} and
\author [Bra] {S.S. Makler}
\address [Arg] {FaMAF, Universidad Nacional de C\'{o}rdoba, 5000 C\'{o}rdoba, Argentina}
\address [Bra] {Instituto de F\'{i}sica,  Universidade do Estado do Rio de Janeiro, and Instituto de F\'{i}sica, Universidade Federal Fluminense, Niter\'{o}i, Brazil.}

\begin{abstract}
We  model the Fock space for the  electronic resonant tunneling through 
a double barrier including the coherent 
effects of the  electron-phonon interaction. The geometry is optimized to achieve
the maximal optical phonon emission  required by a SASER (ultrasound emitter) device.
PACS numbers: 73.20.Dx, 73.40.Gk, 73.50.Rb
\end{abstract}
\end{frontmatter}

The possibility of generating coherent phonons in a double barrier
semiconductor heterostructure was first proposed\cite{BrazJPhys} a few years
ago. This is the basis of a SASER device\cite{NewSci} which transforms the
electric potential energy in a single vibrational mode of the lattice. This
is facilitated by the electronic confinement in a double barrier structure.
The phonon emission appears when the energy of the resonant state is one
quantum $\hbar \omega _{0}$(LO phonon energy) bellow the energy of the
incoming electrons. As in laser devices this is enhanced if the first
excited state of the well lies bellow the Fermi energy and becomes
overpopulated. According to ref.\cite{JPCM98} the emitted LO phonons decay
coherently into a pair of $LO$ and TA phonons the last the useful ones in a
SASER device.

In this paper we want to explore the case in which well's {\it ground} state
mediates the decay of the emitter states into the collector's ones {\it plus}
a phonon. This feature represents a {\it resonance in the electron-phonon
Fock space }and is observed as a satellite peak in the current\cite{Goldman}%
. This resonant condition is tuned directly by the applied voltage and we
expect that its optimization could also provide enough emission of primary
phonons to allow for SASER operation. We carry out the modeling of the
electronic structure and the electron-phonon interaction to get a minimal
structure in the Fock space. Thus, the optimization of the phonon emission
for different geometries of the device (height and width of the barriers,
field intensity) can be discussed in simple terms.

We consider a one-dimensional model for a double barrier including the
interaction with LO phonons in the well, neglecting the effects derived from
the accumulated charge. This will give results comparable to the 3-D case
when $\varepsilon _{F}$ is small, thus limiting the number of traversal
modes; or in the presence of a high magnetic field perpendicular to the
plane of the barriers\cite{Boebinger} which quantize these modes in Landau
levels.We do not consider the phonon-phonon interaction that leads to the
decay of the LO\ phonons.

The Hamiltonian is a sum of an electronic contribution, a phonon
contribution and an electron-phonon interaction term.

\[
{\cal H}={\cal H}_{e}+{\cal H}_{p}+{\cal H}_{e-p} 
\]

\[
{\cal H}_{e}=\sum_{j}E_{j}c_{j}^{+}c_{j}-%
\sum_{j,k}V_{j,k}(c_{j}^{+}c_{k}+c_{k}^{+}c_{j}), 
\]
\[
{\cal H}_{p}=\hbar \omega _{0}\sum_{k[{\rm well}]}b_{k}^{+}b_{k},\,\,\,{\rm %
and\,\,\,\,}{\cal H}_{e-p}=V_{g}\sum_{k[{\rm well}%
]}c_{k}^{+}c_{k}(b_{k}^{+}+b_{k}). 
\]
where $c_{j}^{+}$and $c_{j}$ are electron operators on site $j$, $E_{j}$ is
the site dependent diagonal energy and $V_{j,k}=V\delta _{j\pm 1,k}$ are the
hopping parameters. We assume that the potential drop ${\bf eV}$ is linear
through the double barrier and limited to it. N$_{{\rm L}}$ and N$_{{\rm R}}$
are the number of sites in the left and right barriers and N$_{{\rm w}}$ are
those in the well, the associated lengths are L$_{i}$=N$_{i}$ 2.825\AA .
There is a single well state in the energy range of interest.

Since the most important interaction between electrons and phonons in polar
semiconductors involves longitudinal optical (LO) phonons, only one phonon
mode with frequency \thinspace $\omega _{0}$ is considered. The
electron-phonon interaction is limited to the well region and the coupling
to the phonons is denoted with $V_{g}$. The model is represented
schematically in figure 1.

\begin{figure}[tbp]
\vspace{30mm}
\caption{Each site represents a basis state in the many-body Hilbert space.
The lower row represents localized states in different sites, the sites in
black correspond to the barriers. The upper row represents the same sites
with one phonon. The lines represent the nondiagonal terms in the
Hamiltonian. The potential profile is also shown.}
\end{figure}

For simplicity we restrict the problem to the case in which we have either $0
$ or $1$ phonons with no phonons in the well before the scattering process.
By modifying $V_{g}\rightarrow V_{g}\sqrt{n+1}$ this also represents a
finite temperature emission $n\rightarrow n+1.$ The effective mass is taken
to be $0.067$ m$_{e}$, the LO phonon frequency $\hbar \omega _{0}=36$ meV
and the value of the hopping parameter $V=-7.1018$eV. $V_{g}\,$is taken $10$%
meV which gives a typical electron-phonon interaction strength $%
g=(V_{g}/\hbar \omega _{0})^{2}\ \simeq 0.1$. The barrier heights are $300$%
meV and the Fermi energy $\varepsilon _{F}$ is taken between $10$ and $20$
meV.

This discrete model is solved exactly using a decimation procedure for the
sites in the barriers and the well \cite{Levstein}. The leads are taken into
account by adding a proper self-energy. The transmittances are computed from
the Green's functions for the system \cite{DAmato}.

Let us denote with $T_{0,0}^{R\leftarrow L}$ ($T_{0,0}^{L\leftarrow R}$) and 
$T_{1,0}^{R\leftarrow L}$ ($T_{1,0}^{R\leftarrow L}$) the transmission
coefficients from left (right) to right (left) where the subscripts $0$ and $%
1$ denote the number of phonons in the outgoing (first subscript) and in the
incoming (second subscript) channel. The total current is a sum of an
elastic current $I_{{\rm el}}$and an inelastic current $I_{{\rm in}}$ (with
the emission of one phonon during the scattering process). These currents
can be calculated from the following expressions 
\begin{eqnarray*}
I_{el} &=&(2e/h)\int [T_{0,0}^{R\leftarrow L}f_{L}(\varepsilon
)-T_{0,0}^{L\leftarrow R}f_{R}(\varepsilon )]d\varepsilon , \\
I_{in} &=&(2e/h)\int [T_{1,0}^{R\leftarrow L}f_{L}(\varepsilon
)-T_{1,0}^{L\leftarrow R}f_{R}(\varepsilon )]d\varepsilon ;
\end{eqnarray*}
where $f_{L}(\varepsilon )$ and $f_{R}(\varepsilon )$ are the Fermi
functions for the left and right leads.

\begin{figure}[tbp]
\vspace{30mm}
\caption{Inelastic current as a function of the applied voltage for
different values of NR.}
\end{figure}

For a given configuration of the system a curve of inelastic current vs.
applied bias is obtained and its maximum value $I_{{\rm in}}^{{\rm max}}$
can be extracted. Figure 2\ shows $I_{{\rm in}}$-V curves as we change N$_{%
{\rm R}}$. The peaks in these curves correspond to the inelastic
contributions to the main peak and to the satellite peak in the total
current respectively. This figure also shows that the peaks are shifted to
higher voltages as N$_{{\rm R}}$ is increased. This shows a strong
renormalization of the resonant energies due to the electrodes. In figure 3
we present $I_{{\rm in}}^{{\rm max}}$ vs. N$_{{\rm R}}$ curves for different
values of N$_{{\rm L}}$. These curves exhibit a maximum for $I_{{\rm in}}^{%
{\rm max}}$ as a function of N$_{{\rm R}}$. The optimal configurations
correspond to asymmetric structures with wider right barriers. This can be
understood by means of the following argument. Increasing the lifetime of
the electrons in the well favors the electron-phonon interaction and thus
increases the inelastic current. This can be done by choosing wider (or
higher) barriers. In spite of this, as an effect of the asymmetry produced
by the applied bias, the lifetime is still controlled mainly by the right
barrier. On the other hand increasing the length of the barriers increases
the reflectivity of the device diminishing the currents, here it is the left
barrier which plays the main role. Then there is a trade off between these
two effects that maximizes the phonon emission.

\begin{figure}[tbp]
\vspace{30mm}
\caption{Maximum inelastic current as a function of NR for Nw=20 and Ef=20
meV.}
\end{figure}

In summary, we have used a simple model to show that the asymmetry in double
barrier structures plays an important role in the $I_{{\rm in}}$-V
characteristics and to predict how it can be controlled to optimize LO\
phonon emission. In particular we show that the optimal configuration
corresponds to a collector barrier with a length which doubles that of the
emitter.

We acknowledge finantial support from CONICET, SeCyT-U.N.C., AnPCyT and an
international grant from Andes-Vitae-Antorchas.

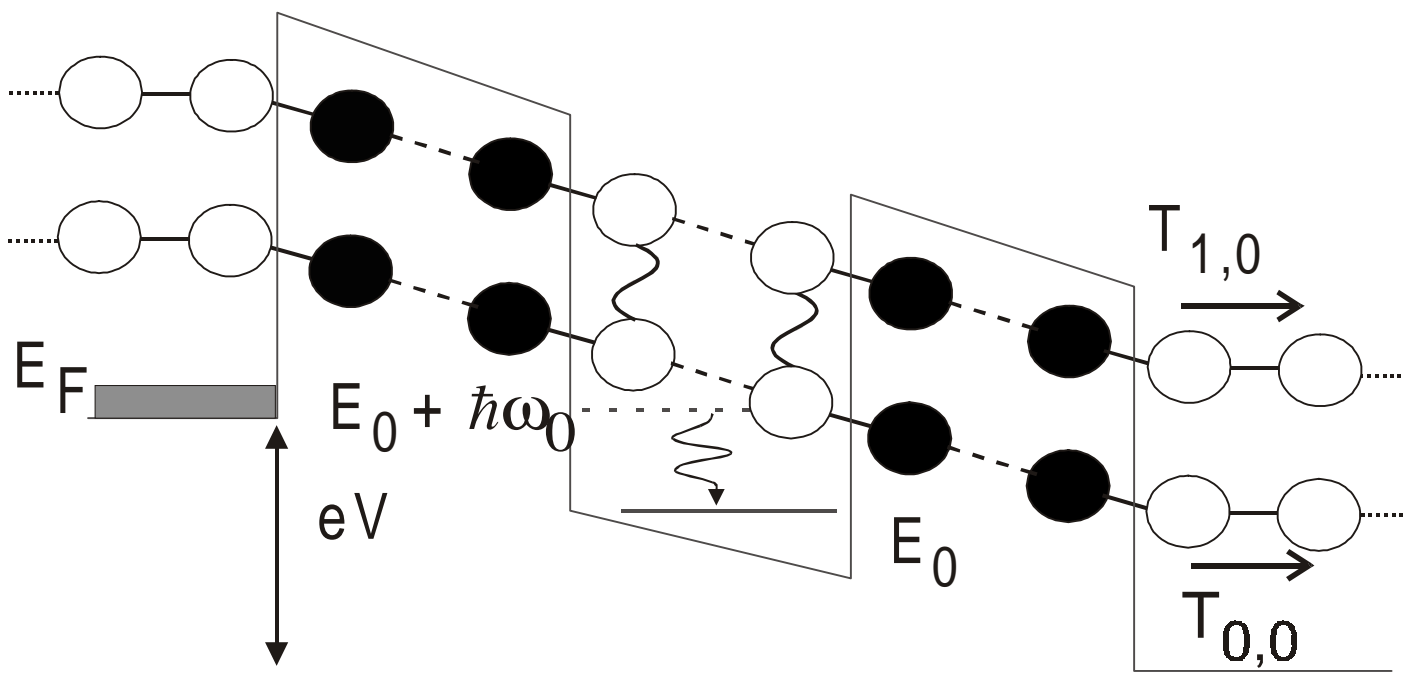
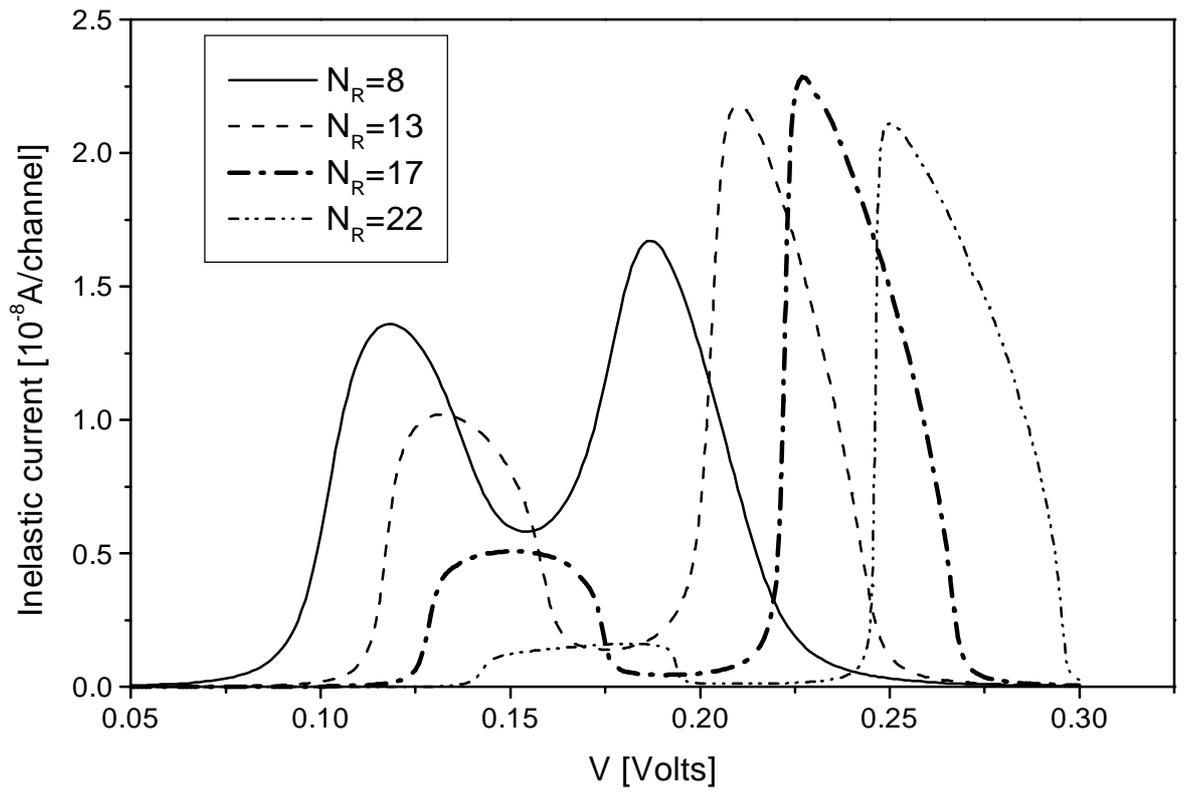
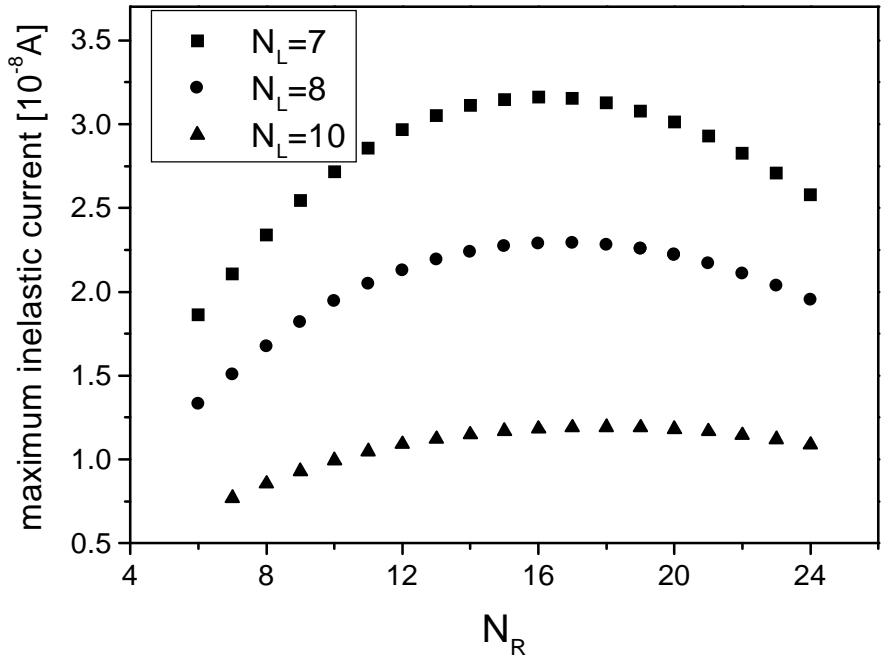
\end{document}